\documentclass{optica-article}

\journal{opticajournal} 

\articletype{Research Article}

\usepackage{lineno}

\begin{document}

\title{Focused axisymmetric spatially chirped beams}

\author{E. C. Nelson,\authormark{1,*} K. J. Charbonnet,\authormark{1} H. H. Effarah,\authormark{1} T. Reutershan,\authormark{1} K. D. Chesnut,\authormark{1} and C. P. J. Barty\authormark{1}}

\address{\authormark{1}Department of Physics and Astronomy, University of California, Irvine, CA 92697}

\email{\authormark{*}ecnelson@uci.edu} 


\begin{abstract*} 
A characterization of the focused space-time structures of radially chirped beams is provided, detailing different tunable properties such as: variable on-axis centroid velocity, symmetric pulse front tilt, transverse intensity modulations, and polarization states. 
While the practical generation of ideal radially chirped beams and polarizations can be problematic, it is shown that the primary characteristics of these beams can be mimicked with simple arrays of axisymmetric, 1D spatially chirped beams. 

\end{abstract*}

\section{Introduction}
Controlling the space-time structure of ultrashort pulses has become an active area of interest for generating exotic structures of focused light \cite{space-time-structure}.
One technique originating from the concept of chirped beam amplification \cite{barty-patent}, known as simultaneous spatial and temporal focusing (SSTF) \cite{durst,durfee_abcd,archer}, results in a focused light field that exhibits a pulse front tilt (PFT) with diffraction-limited spot size and transform-limited pulse duration. 
These properties have enabled electron beam steering \cite{wilhelm_thesis,wilhelm, Hunt}, enhanced axial resolution in microscopy \cite{Oron,durst,Papa}, and micromachining \cite{Vitek, kammel, Patel}.
The space-time couplings required for SSTF can be fabricated through spatial light modulators \cite{4dlight,yessenov} and with dispersive or diffractive optics \cite{akturk,song}.
Typically, a single-pass, two-grating compressor is used to generate a spatially chirped beam in one dimension that is subsequently focused, generating the SSTF space-time structure \cite{durfee_abcd,block2,He,Zhang}.

In this manuscript, the axisymmetric extension of 1D spatial chirp, referred to as radial spatial chirp, is evaluated and the space-time structure of the focal intensity for various focusing conditions and polarization states is presented.
In principle, radially chirped beams can be created using circular gratings \cite{circular_grating}, also known as concentric-ring gratings \cite{ghadyani2011concentric, Lanier, zhou2019concentric}, in a similar single-pass, two-grating configuration.
This results in a radially dependent spatial chirp from the diffraction, shown in Fig. \ref{fig:chirp_polarizations}.
The radial spatial chirp maintains many of the qualities of SSTF pulses with the added symmetrization of the pulse front and the enabled capacity for vector polarization states. 
Figure \ref{fig:chirp_polarizations} shows all the possible variations of radial chirp and polarization configurations compared to a 1D spatial chirp.

Radially chirped beams have been preliminarily studied to various extents.
The focus of a radial spatial chirp generated with an axicon pair, which provides a positive spatial chirp, has been studied for its bessel-like space-time structure \cite{Clerici}.
Radially chirped picosecond pulses have been examined for the generation of non-differentiable angular dispersion \cite{yessenov} to aid in propagation invariant space-time pulses with variable centroid velocity tuning \cite{Froula,Liberman}.
The nonlinear intra-material effects of a focused radially chirped beam have been numerically studied by varying the initial temporal group delay dispersion of the chirped pulse \cite{Lanier}.
The variation in longitudinal, on-axis intensity profiles has also been studied \cite{archer_thesis}.
Additionally, focused linearly chirped, radially polarized beams have been examined \cite{Jolly_radial} as a unique form of vector beams with spatio-temporal couplings.

Radial spatial chirp can be approximated through the coherent combination of appropriately spaced and oriented 1D spatially chirped beams.
This allows for the use of current grating technology to generate beams comparable to radial spatial chirp with high damage threshold optics that have been optimized for high power lasers to enable the use of these space-time vector beams in intense laser-matter interactions, where high fluence values limit the use of transmissive optics and polarizers.
In this manuscript, a comprehensive characterization of the properties of radially chirped beams and their approximations is provided.


\section{Methods}


\begin{figure}[ht!]
\centering\includegraphics[width=\linewidth]{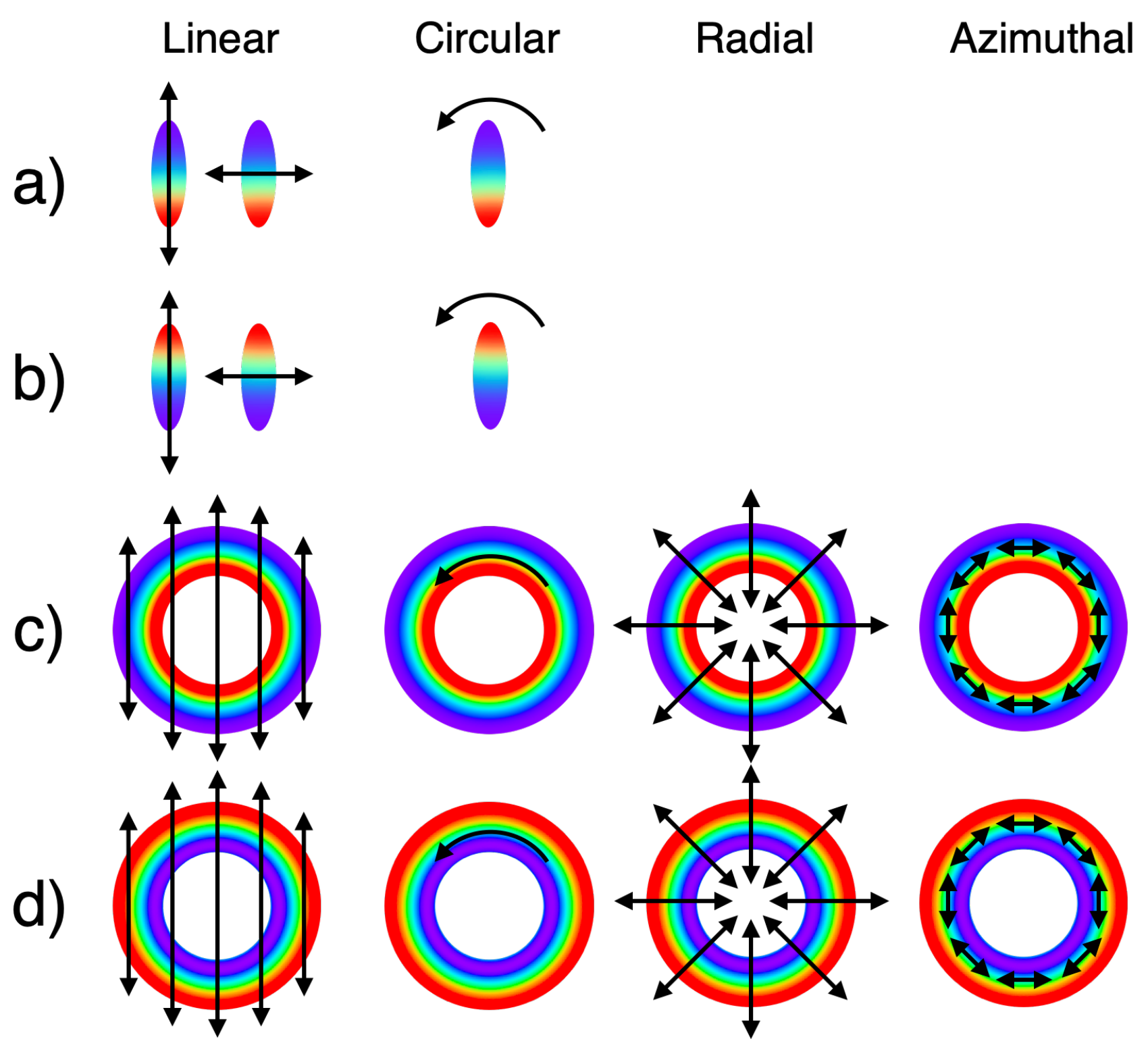}
\caption{The various spatial chirp patterns and polarization orientations for (a),(b) 1D chirp and (c),(d) 2D chirp. The color represents the spatial dependence of the spectrum at the output of a single-pass, two grating compressor and the black arrows represent polarization direction.
}
\label{fig:chirp_polarizations}
\end{figure}

\begin{figure}[t!]
\centering\includegraphics[width=\linewidth]{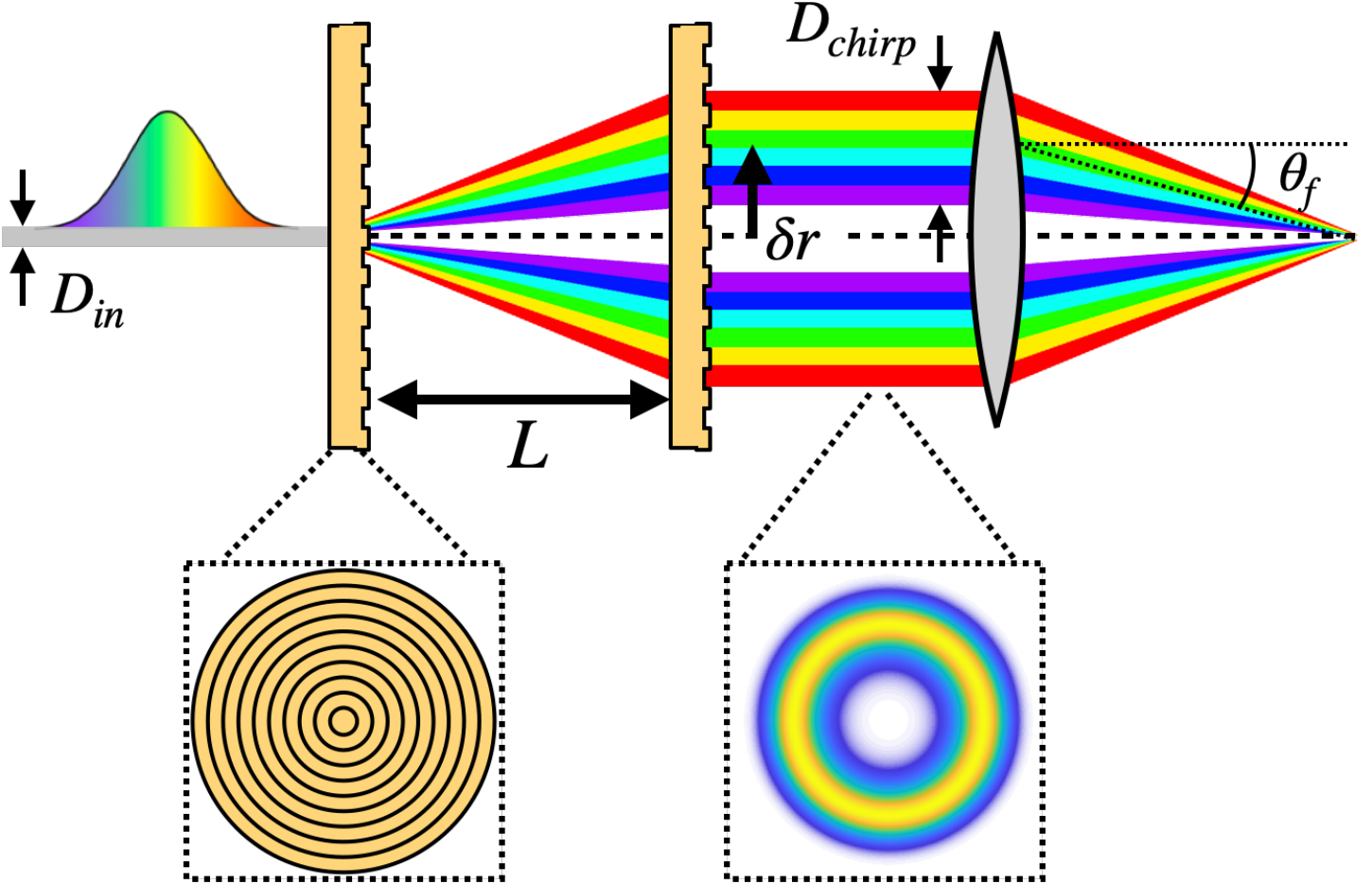}
\caption{A single pass, concentric ring grating pair used to generate radial spatial chirp. The insets show the transverse profile of a concentric ring grating and the annular beam profile after the second grating.}
\label{fig:concentric_grating}
\end{figure}

A pair of concentric ring gratings (Fig. \ref{fig:concentric_grating}) can be used to generate a negative radial spatial chirp pattern, where the lower frequencies are diffracted farther from the propagation axis, shown in Fig. \ref{fig:chirp_polarizations}(d).
The amount of spatial chirp is quantified by the beam aspect ratio
\begin{equation}
    \beta_{\text{BAR}}=\frac{D_\text{chirp}}{D_\text{in}},
\label{eq:BAR}
\end{equation}
where $\beta_{\text{BAR}}$ is defined as the ratio of the effective chirped beam diameter, $D_{\text{chirp}}$, to the initial un-chirped beam diameter, $D_{\text{in}}$.
The distance of the center carrier frequency of the pulse from the propagation axis is defined as $\delta r$.
In this configuration, $D_{\text{chirp}}$ and $\delta r$ are coupled based on the grating equation, where increasing the grating separation increases $D_{\text{chirp}}$ and $\delta r$.
Positive spatial chirp, where the the longer wavelengths are closer to the propagation axis (Fig. \ref{fig:chirp_polarizations}(c)), is more difficult to achieve.
A reflective axicon pair may be used to invert the orientation of the spatial chirp from a grating pair and also allow an independent tuning of $\delta r$ from changing the distance between axicons.
A transmissive, refractive axicon pair can also be used to generate positive spatial chirp \cite{Clerici}, where shorter wavelengths refract more through the axicon material, placing them on the outside edge of the beam. 
In order to achieve large amounts of spatial chirp over reasonable propagation distances, axicons with apex angles near the critical, total internal reflection angle must be used in order to increase the angular dispersion from the refraction. 
Additionally, by utilizing ring lenses \cite{RL} the spatial chirp from a grating pair can be relay imaged to flip the sign of spatial chirp. 
Using additional transmissive optics is detrimental to the propagation of ultrashort pulses, and must be pre-compensated in the system.
Since the axicons and rings lenses are relatively thick, high order dispersion terms may not be fully compensated for, distorting the temporal shape of the pulse.
Using refractive axicons can also introduce spatial modulations in the transverse beam profile due to clipping on the apex of the axicon \cite{axicon_clipping}.

\begin{figure}[h!]
\centering\includegraphics[width=\linewidth]{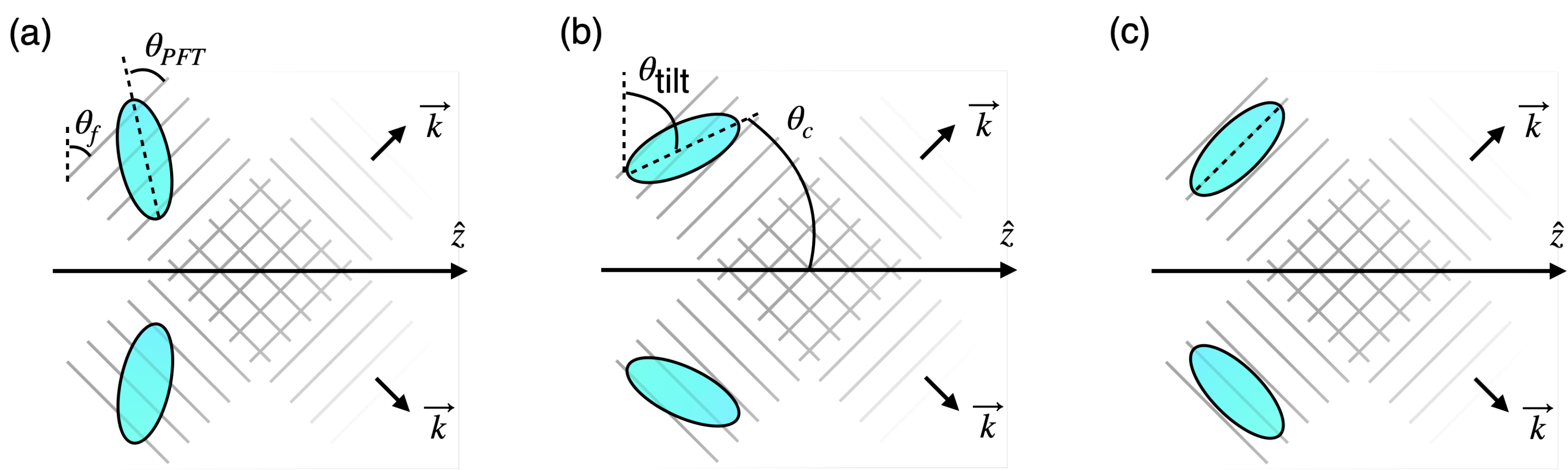}
\caption{Schematic of different focusing regimes depending on the sign of spatial chirp. The pulse front tilt angle, $\theta_{\text{PFT}}$ is defined with respect to the phase fronts which are shown as solid lines. $\theta_f$ is the angle of the phase fronts from the off-axis focusing. (a) negative PFT, (b) positive PFT, (c) no PFT.}
\label{fig:vg_explain}
\end{figure}
The orientation of the PFT generated from the focused spatially chirped beams relative to the propagation axis changes for different chirps.
In a focused, spatially chirped beam, the higher frequency portion of the spatial chirp leads in time through the focus \cite{durfee_abcd}, therefore the sign of the chirp changes the propagation effects throughout the focus.
Fig.~\ref{fig:vg_explain} represents a schematic of the focusing geometry for the symmetric beams near the focus.
The phase fronts are illustrated as straight lines, and are tilted by an angle $\theta_f$ due to the off-axis nature of the initial annular profile when interacting with a focusing optic.
The angle the pulse front tilt makes with respect to the phase fronts due to the phase delay from focusing a spatial chirp is defined as $\theta_{\text{PFT}}$.
Fig.~\ref{fig:vg_explain}(a) shows a cross section of the symmetric beam near the focus of a negative spatial chirp (Fig.~\ref{fig:chirp_polarizations}(d)), where the center of the beam always leads the pulse front.
The positive spatial chirp (Fig.~\ref{fig:chirp_polarizations}(c)) focuses as Fig.~\ref{fig:vg_explain}(b).
For reference, an annular beam with no chirp, and therefore no PFT, focuses as Fig.~\ref{fig:vg_explain}(c).
The total effective tilt from the contribution of these two effects at the focus is defined as:
\begin{equation}
    \tan \theta_{\text{tilt}}= \tan \theta_f + \tan \theta_{\text{PFT}}.
\label{eq:tilt}
\end{equation}
with
\begin{equation}
    \tan \theta_{f}=\frac{\delta r}{f}
\label{eq:focusing}
\end{equation}
and
\begin{equation}
    \tan \theta_{\text{PFT}}=\frac{w_{\text{in}}\omega_0 \sqrt{\beta_{\text{BAR}}^2-1}}{\Delta \omega f}
\label{eq:PFT}
\end{equation}
where $w_{\text{in}}$ is the $1/e^2$ radius of the input spot intensity, $\omega_0$ is the center frequency of the pulse, and $\Delta \omega$ is the $1/e^2$ half-width bandwidth of the spectral intensity.
The half-cone angle is defined as $\theta_c=\pi/2-\theta_{\text{tilt}}$.

While the creation and tuning of radially chirped beams is challenging due to the unconventional optics needed, it will be shown that an approximate form of the radial spatial chirp can be achieved through a multi-beam superposition of properly oriented 1D spatial chirps.
An advantage to this approach is the ability to create symmetric spatial chirps of ultrahigh intensity laser pulses, as high damage threshold optics are used in the multi-beam approach.
Instead of using custom transmissive concentric ring gratings, conventional optics utilizing current grating technology can be used to generate the spatial chirp using highly efficient grating groove densities and angles of incidence (AOI).
Additionally, this configuration also provides independent control of $\delta r$, the polarization, and the sign of spatial chirp through simply modifying the orientation of each singular beam.

The spatial chirps in the following simulations are derived from the grating equation to ensure accurate space-time profiles when propagated \cite{Nelson}.
The fields will initially be analyzed from the output of a single-pass, concentric ring grating pair assuming uniform diffraction efficiency as a function of beam polarization.
The temporal chirp from the grating pair is also assumed to be pre-compensated such that there is no residual temporal dispersion at the output of the grating pair.
An analysis of the properties of the space-time structure assuming $\beta_{\text{BAR}}$, $\delta r$, and chirp orientation can be independently controlled follows.
Finally, a proposed multi-beam scheme will be presented, which enables practical, independent tuning of the variables while maintaining a close approximation to the full radial field.
The spatially chirped beams are propagated through space using the vectorial Rayleigh-Sommerfeld \cite{sommerfeld,wolf,goodman, voelz} convolution technique, along with the Bluestein method \cite{bluestein_propagation,bluestein_og}, which allows for an arbitrarily sampled input plane and propagation plane for non-paraxial focused pulses using the provided Python3 software \cite{code}.
The simulated laser pulse is based on a Ti:Sapphire oscillator, with a center wavelength of 800~nm, a full-width-at-half-maximum Gaussian bandwidth of 100~nm, and a $1/e^2$-intensity beam radius of 0.5~mm.
Various configurations of spatial chirp, $\delta r$, and polarization are modeled and propagated through a focal volume. 
In each configuration, the pulse is normalized to have 1~J of energy.
The fields are propagated in the spatio-spectral domain to any longitudinal position within the focus, and are Fourier transformed to visualize the space-time structure. 

\section{Results}
\begin{figure}[h!]
\centering\includegraphics[width=\linewidth]{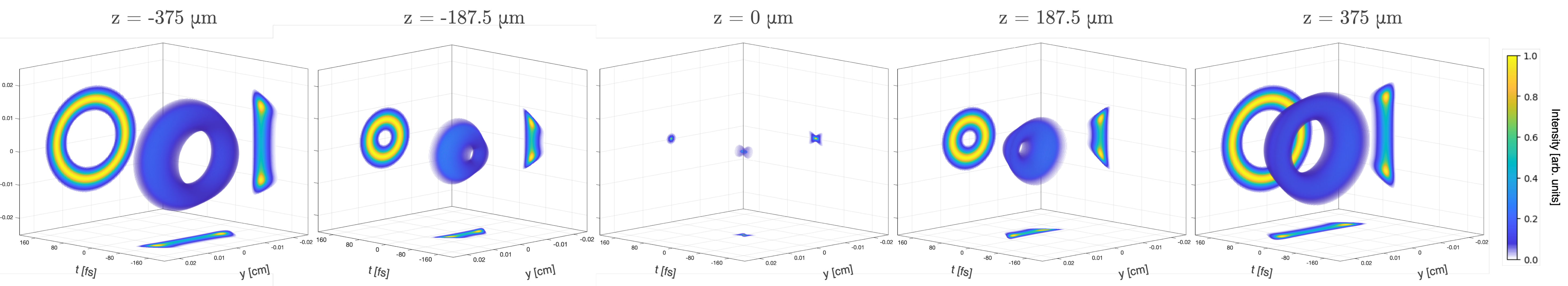}\caption{The propagation of a focused radially chirped beam. The negative radial chirp was generated with a grating pair of 700 lines/mm at $0^o$ AOI, $L=3$ cm and focused with an ideal lens of focal length $f=5$ cm. Each longitudinal position has the maximum intensity normalized to a value of 1 to better highlight the structure of the field as it propagates.}
\label{fig:propagation}
\end{figure}

The natural, focused output of a concentric ring grating pair is simulated for various grating separations.
The only variables that may be changed which affect the focal volume structure are the grating separation, groove density, and focal length.
The gratings are assumed to have a groove density of 700 lines/mm and the focusing optic is treated as the phase from an ideal lens with focal length $f=5$ cm. 
The pulse is assumed to be linearly polarized in the x-dimension and transform-limited after the lens.
First, the space-time structure of the pulse as it propagates through the focus is visualized in Fig. \ref{fig:propagation}.
The chirp pattern from the grating pair before focusing is that of Fig. \ref{fig:chirp_polarizations}(d).
In this arrangement, the high frequency content of the spatial chirp is on the inside portion of the beam, and it is this edge that leads in time \cite{durfee_abcd}, resulting in the conical structure displayed in Fig. \ref{fig:propagation}.
The apex of the conical structure leads along the axis as it approaches the focus, which then flips in orientation after the focus.
\begin{figure}[h!]
\centering\includegraphics[width=\linewidth]{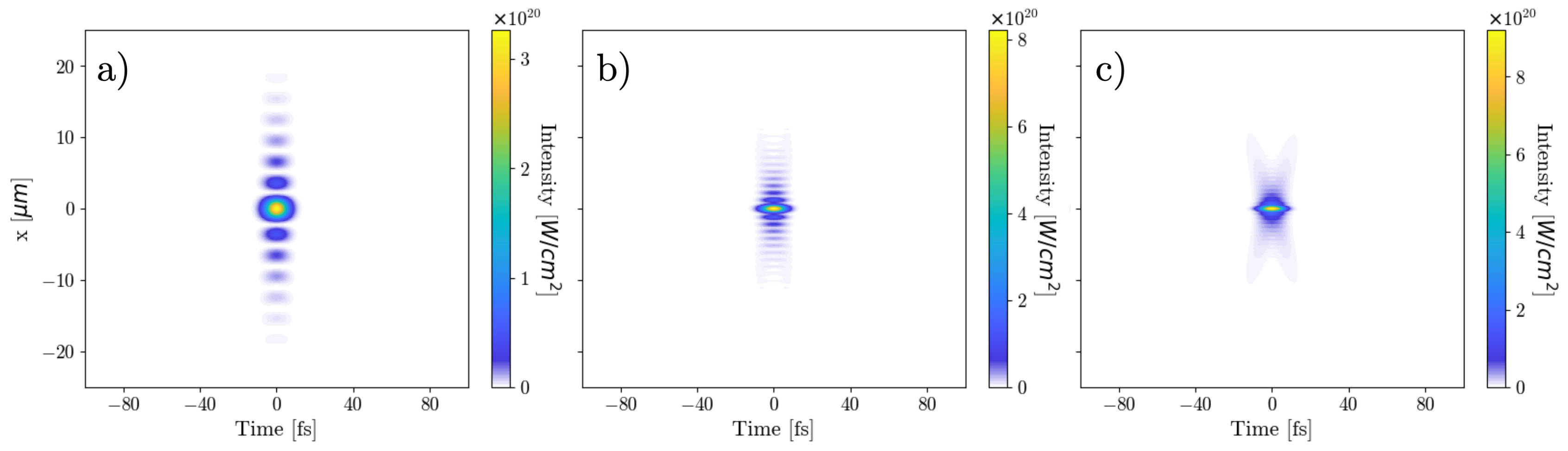}\caption{(x,t) representations of the transverse intensity of the focus ($f=5\; \text{cm})$ for a linearly polarized pulse spatially chirped with a $700$ lines/mm grating pair. The plots (a-c) represent pulses with grating separations of 1 cm, 3 cm, and 5 cm, respectively.}
\label{fig:focus_variation_nat}
\end{figure}
At the focal plane, the pulse exhibits an x-shaped profile, which is analyzed more closely in Fig. \ref{fig:focus_variation_nat}, which shows the space-time structure of the total intensity ($E_x^2+E_y^2+E_z^2)$ as a slice in (x,t) space at y=0 for three cases of grating separation.
As grating separation increases, $\beta_{\text{BAR}}$ and $\delta r$ increase, resulting in the space-time structure becoming more x-shaped, having more transverse intensity modulations, and an increase in peak intensity.
The modulations begin to disappear Fig. \ref{fig:focus_variation_nat}(c), as the f-number of the system becomes small enough that a portion of the pulse becomes longitudinally polarized and contributes more significantly to the spatial field profile.
Due to the coupled nature of the grating pair, it is not easy to tune these parameters independently.
The methods described in the previous section offer a potential avenue for creating arbitrary radial, spatially chirped beams.
Independent control of both $\beta_{\text{BAR}}$ and $\delta r$ offers the ability to shape the space-time profile of the pulse, modifying both the transverse intensity modulations, the conical angle as it focuses, and the amount of temporal focusing.
\begin{figure}[h!]
\centering\includegraphics[width=\linewidth]{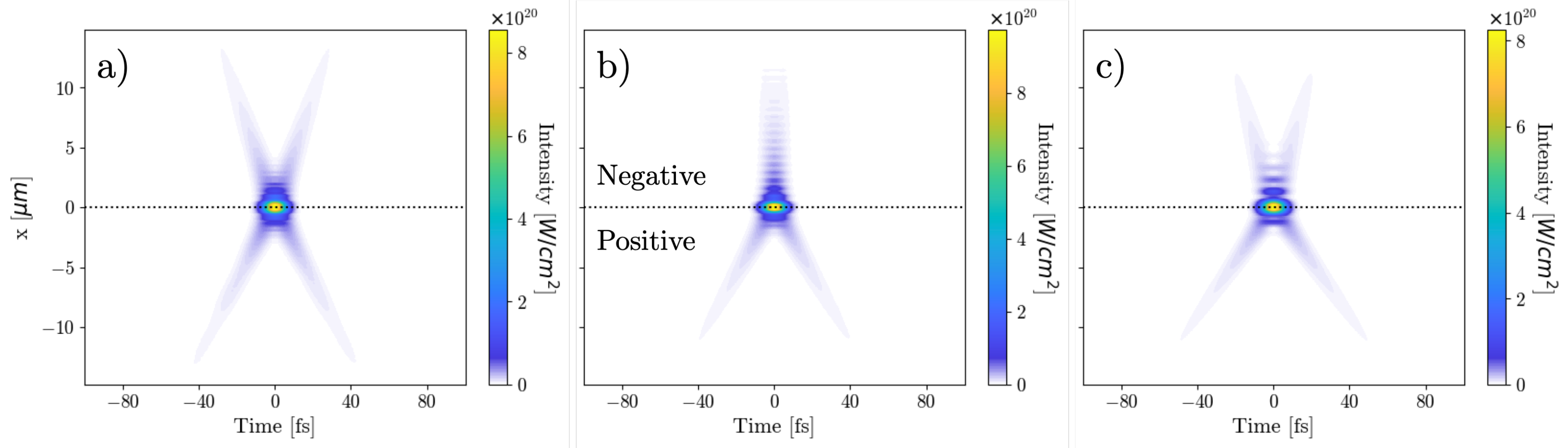}\caption{(x,t) representations of the transverse intensity of the focus ($f=5\; \text{cm})$ for a linearly polarized pulse spatially chirped with a $500$ lines/mm grating pair. The plots (a-c) represent pulses with $\delta r=4$ cm and $\beta_{\text{BAR}}=2$, $\delta r=3$ cm and $\beta_{\text{BAR}}=5$, $\delta r=2$ cm and $\beta_{\text{BAR}}=10$, respectively. Within each plot, negative spatial chirp is plotted in $x>0$ and positive spatial chirp is plotted in $x<0$ to provide direct visual comparison. A horizontal dotted line at $x=0$ in each plot indicates this separation.}
\label{fig:focus_variation}
\end{figure}
Fig. \ref{fig:focus_variation} shows the space-time profiles of the focal plane for six cases, where the sign of spatial chirp, $\beta_{\text{BAR}}$, and $\delta r$ are assumed to be independently tunable and are varied.
The pulse shapes are able to be modified much more drastically than the constrained case shown in Fig. \ref{fig:focus_variation_nat}.
Through changing $\delta r$, less transverse intensity modulation is observed for smaller values.
The cone angle is also tuned more drastically, as the tilt angle depends on contributions from $\delta r$ and $\beta_{\text{BAR}}$ in Eq. \ref{eq:tilt}.
The negatively chirped case has much more variety in this cone angle than the positive chirp. 
\begin{figure}[b!]
\centering\includegraphics[width=\linewidth]{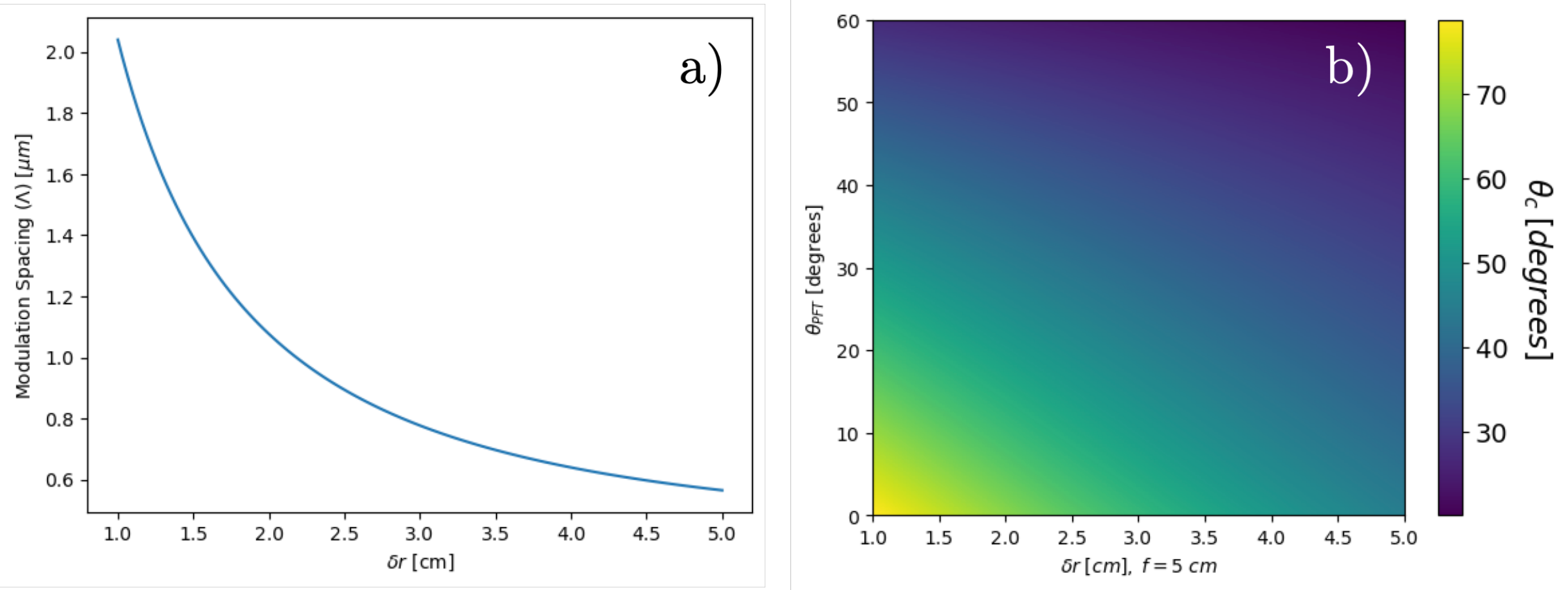}\caption{(a) The spacing between fringes of the focused radially chirped beam as a function of $\delta r$. (b) The cone angle for the radially chirped beam for different values of $\theta_f$ and $\theta_{\text{PFT}}$.}
\label{fig:trends}
\end{figure}
The negative chirp imparts a negative sign in $\theta_{\text{PFT}}$ from Eq. \ref{eq:tilt}, which counteracts $\theta_f$ from the geometric focusing, invoking an interplay between the two angle contributions.
In Fig. \ref{fig:focus_variation}(a), $\theta_{\text{PFT}}$ is negative and larger than $\theta_f$ and contributes a more significant portion to $\theta_{\text{tilt}}$.
In Fig. \ref{fig:focus_variation}(b), these angles become approximately equal to each other and the apparent cone angle is $\theta_c=90^o$, in an effect known as PFT matching.
This configuration maximizes the transverse spatial overlap (and number of interference fringes) at the focus, which has been shown to enhance interference fringe contrast in autocorrelation measurements\cite{meier}.
With Fig. \ref{fig:focus_variation}(c), $\theta_{\text{PFT}}$ is still negative but smaller than $\theta_f$, and so it is the focusing angle that dominates the apparent tilt.
The positive chirp in all three configurations result in $\theta_{\text{PFT}}$ and $\theta_f$ adding together.
Only six cases are shown for linear polarization, additional figures can be found in Fig. S1 in Supplement 1.

Fig. \ref{fig:trends}(a) shows the trends of the transverse intensity modulations.
The spacing of the transverse intensity modulations, $\Lambda$, follows the form \cite{interference}:
\begin{equation}
    \Lambda=\frac{\lambda_0}{2 \sin \theta_f},
\label{eq:fringe}
\end{equation}
which is the equation for two plane waves interfering at an angle $\theta_f$ from the optical axis, with the central wavelength, $\lambda_0$.
At the focus, the radially chirped beam exhibits this same interference pattern, albeit with radial symmetry.
The interplay between $\theta_{\text{PFT}}$ and $\theta_f$ is shown in Fig. \ref{fig:trends}(b), where the cone angle is plotted as a function of these two variables.
The cone angle of the pulse structure can be tuned through varying either of the variables.
Modifying $\delta r$ (and thus $\theta_f$) results in both a modification of the pulse front cone and the transverse intensity modulations.
Changing $\theta_{\text{PFT}}$ only changes the cone angle at the focal plane.
Varying $\theta_{\text{PFT}}$ through $\beta_{\text{BAR}}$ modifies the pulse as it propagates, as this value changes the temporal focusing due to the spatial chirp spreading out the frequency content and decreasing local pulse duration with the form \cite{durfee_abcd}:
\begin{equation}
    \Delta t (z)=\Delta t_0 \sqrt{\frac{1+\beta_{\text{BAR}}^2 \left(z/z_R\right)^2}{1+\left(z/z_R\right)^2}},
\label{eq:pulse_duration}
\end{equation}
where $\Delta t (z)$ is the local pulse duration at a position z (z=0 is at the focus), $\Delta t_0$ is the original transform limited pulse duration, and $z_R=\pi w_0^2/\lambda_0$ is the Rayleigh length.
\begin{figure}[b!]
\centering\includegraphics[width=\linewidth]{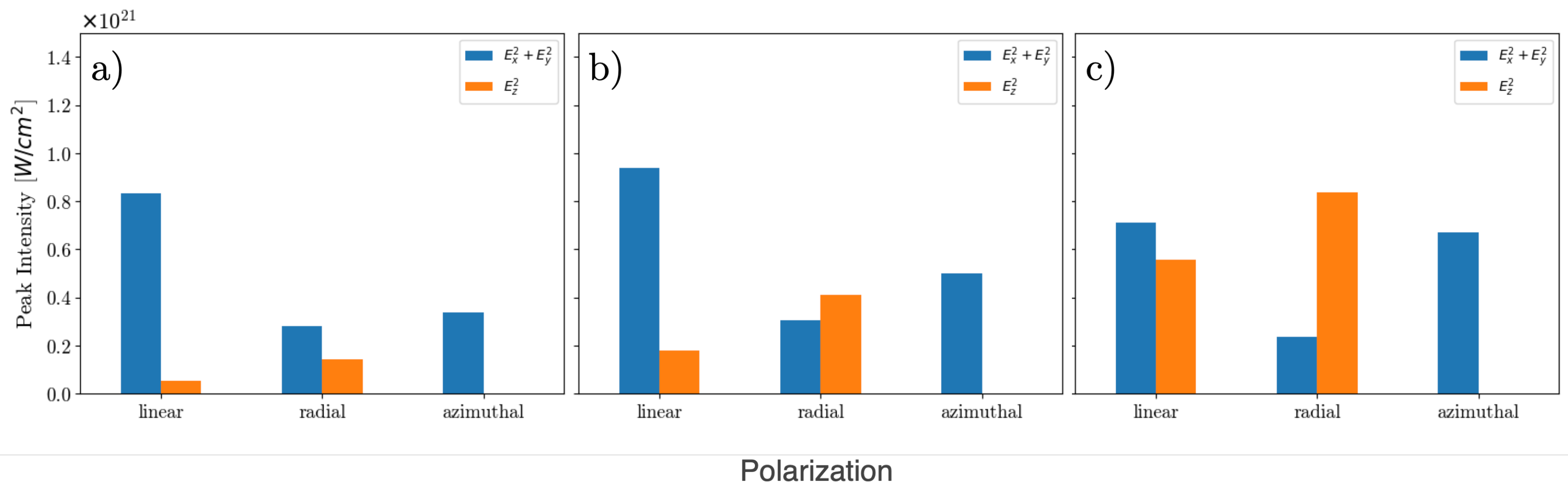}\caption{The peak intensities for the transverse and longitudinal fields for linear, radial, and azimuthal polarizations with (a) $\delta r= 2$ cm, (b) $\delta r= 3$ cm, and $\delta r= 4$ cm. The radially chirped beams were generated with a grating pair of 500 lines/mm and $f=5$ cm.}
\label{fig:pol_trend}
\end{figure}
The annular nature of the radially chirped pulse also leads to the vector polarization states illustrated in Fig. \ref{fig:chirp_polarizations}.
The polarization states change the ratio of the transverse field components to the longitudinal field component of the pulse at the focal plane. 
Radially polarized pulses have been shown to generate large longitudinal field components, while azimuthal polarizations minimize longitudinal field.
The peak intensity of the transverse field ($E_x^2+E_y^2$) and the peak intensity of the longitudinal field ($E_z^2$) is shown in Fig. \ref{fig:pol_trend}.
The (x,t) space-time profiles for radial and azimuthal polarizations are shown in Figs. S2 and S3 of Supplement 1.
As the radially polarized fields come to a focus, the on-axis component of the transverse field evolves into a longitudinal field along the propagation direction.
For azimuthal polarization, the transverse spatial distribution of the transverse fields match that of radial polarization, but no longitudinal field is generated, resulting in higher peak transverse intensities.
The longitudinal field can become quite significant for low f-number systems and can achieve sub-diffraction-limited spot sizes \cite{radial_spot1,radial_spot2}.

\begin{figure}[b!]
\centering\includegraphics[width=\linewidth]{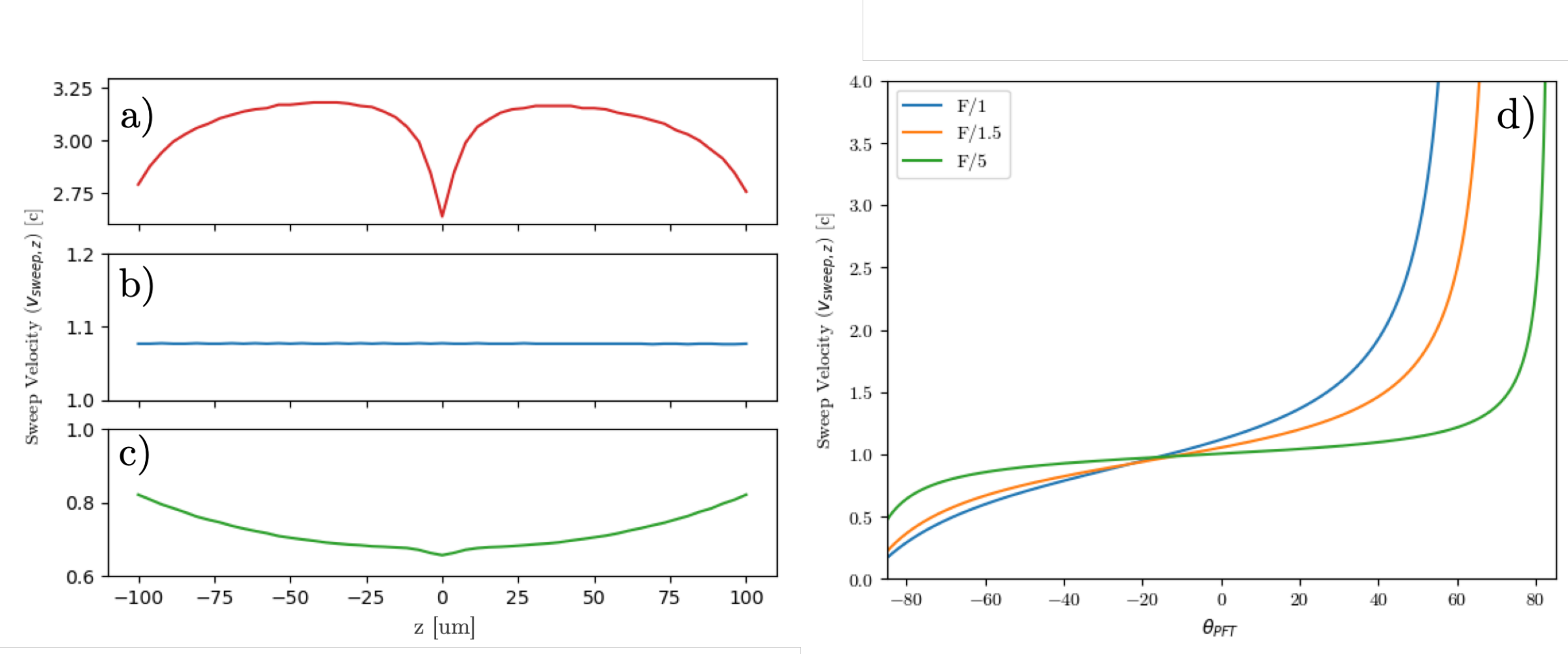}
\caption{Left: The centroid velocity as a function of position from the focus of a (a) positive spatial chirp, (b) no spatial chirp, and a (c) negative spatial chirp. The simulation used $\beta_{\text{BAR}}=20$, $f=5\; \text{cm}$ generated with 500 lines/mm groove density gratings. (d) The on-axis centroid velocity at $z=0$ as a function of PFT angle for different f-number focusing systems by varying focal length with $\delta r = 0.5$ cm.}
\label{fig:group_velocity}
\end{figure}

The tunability of the PFT angle of the radial chirp allows for more convenient control of the on-axis centroid velocity compared to that of an annular beam focused through an axicon \cite{conical}.
By tuning the grating separation or groove density to give varying values of PFT and $\delta r$, in conjunction with different focusing conditions, the sweep speed of the peak intensity centroid along the optical axis can be tuned from sub-luminal to super-luminal values.
The sweep velocity for these radial, spatially chirped beams is not constant throughout propagation. 
Fig. \ref{fig:group_velocity}(a)-(c) shows how the sweep velocity changes during propagation near the focus for different values of spatial chirp. 
The variable centroid velocity is due to two primary effects. 
First, the PFT angle of a focused, spatially chirped beam does not remain constant throughout propagation and gradually increases to a maximum at the focus, following the form \cite{wilhelm_thesis}:
\begin{equation}
    \theta_{PFT}=\tan^{-1}\left[ \frac{w_{in}\sqrt{\beta_{BAR}^2-1}}{\Delta \omega f} \left( \frac{1}{1+z^2/z_R^2} \right) \right]
\label{eq:pft_prop}
\end{equation}
which is a more general form of Eq. \ref{eq:focusing} to account for z-position.
For positive PFT, the increase in PFT angle during propagation towards the focus increases the sweep velocity. 
Similarly, the decrease in PFT angle during propagation away from the focus decreases the sweep velocity. 
The opposite is true for negative PFT. 
The second effect is that the centroids of a focused spatially chirped beam do not propagate in a straight line from the lens to the focus, but rather translate across the transverse dimension \cite{Nelson}. 
This shift causes the centroid velocity to rapidly slow near the focus for both negative and positive spatial chirps. 
The combination of these effects results in the opposite concavity of centroid velocity for negative and positive spatial chirp but with similar dips near the focus.

The longitudinal centroid velocity at the focus $(z=0)$ follows the form:

\begin{equation}
    v_{\text{sweep},z} = \frac{c}{n \cos \theta_f} \left[ 1- \; \tan \theta_f \frac{w_{in} \sqrt{\beta_{BAR}^2-1}}{\Delta \omega f \left( 1 + \theta_f
    ^2 \right)} \right]^{-1}
\label{eq:vg}
\end{equation}
where $n$ is the material index of refraction of the medium the pulse is propagating in (n=1 for vacuum propagation).
Fig. \ref{fig:group_velocity}(d) shows how the on-axis centroid velocity at $x=y=z=0$ varies for three different f-number systems, where $F/\#=f/(2\delta r)$. 
The on-axis sweep speeds reach an asymptote for large PFT angles, and asymptote earlier for lower $F/\#$ systems (or larger $\theta_f$).
Through careful tuning of the system, large super-luminal and sub-luminal velocities may be achieved.

\subsection{Multi-Beam Approach}
Generating a radial chirp with a concentric ring grating pair is a conceptually simple method, but has limited tunability, requiring off-axis reflective axicons or other specialty optics to independently control $\beta_{\text{BAR}}$, $\delta r$, and chirp orientation.
For  large values of $\beta_{\text{BAR}}$ and $\delta r$, large aperture concentric ring gratings and focusing optics would be needed to match the full radial beam size.
For example, the case considered with $\beta_{\text{BAR}}=10$ and $\delta r=4$ cm requires a second concentric ring grating approximately 4 inches in diameter.
Additionally, the concentric ring grating introduces polarization inefficiencies across the beam.
\begin{figure}[h!]
\centering\includegraphics[width=0.8\linewidth]{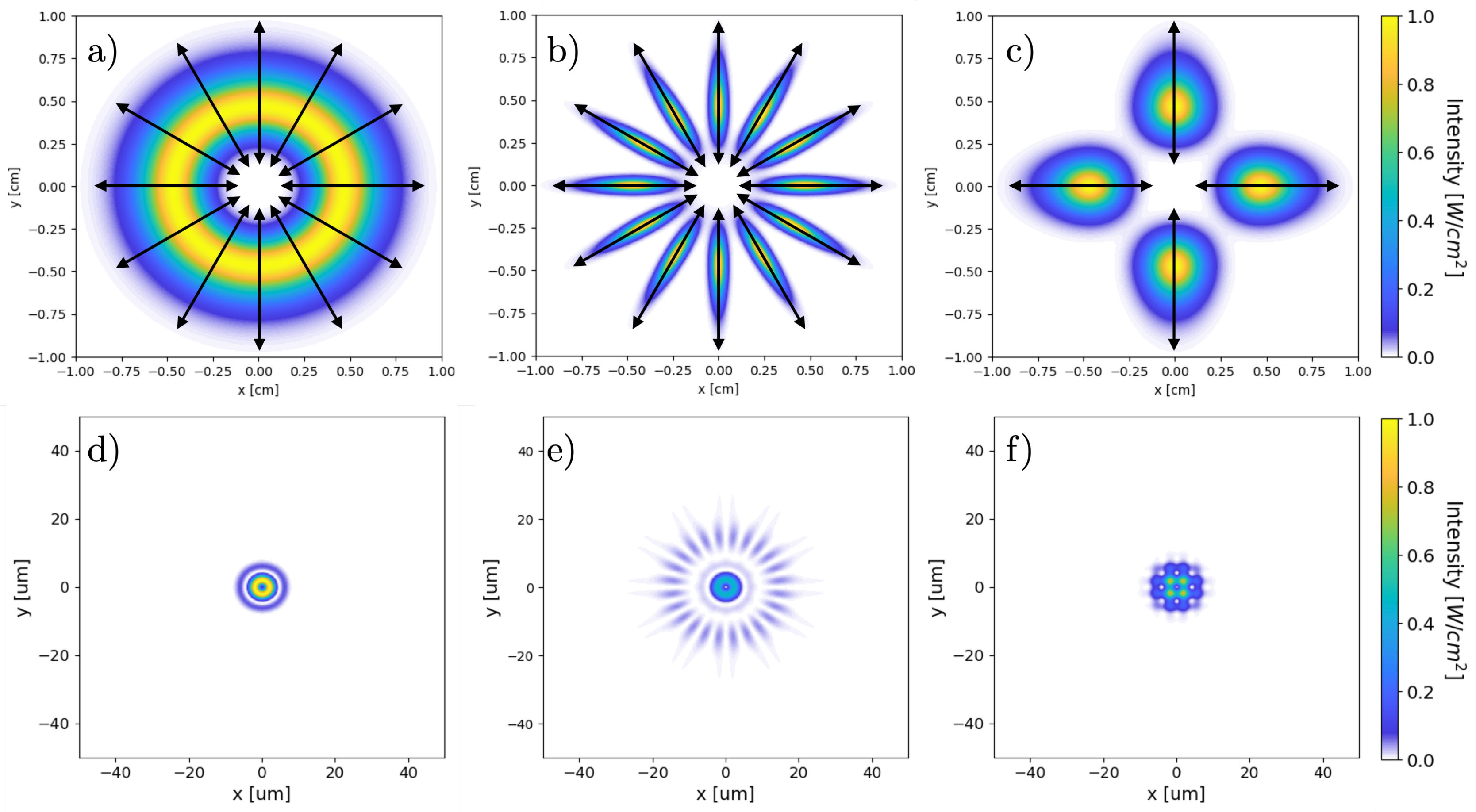}
\caption{The temporally integrated transverse (x,y) input plane of (a) a radially chirped beam, (b) an equivalent twelve beam approximation, and (c) four beam approximation. The transverse (x,y) focal plane at t=0 of (a) radially chirped beam, (b) an equivalent twelve beam approximation, and (c) equivalent four beam approximation. The beam is modeled with an effective $\beta_{\text{BAR}}=5$, generated with a grating groove density of 500 lines/mm, and $\delta r=0.5$ cm}
\label{fig:beam_approx}
\end{figure}

A method of approximating a radial chirp is shown using multiple beam lines that are coherently combined, which offers the benefit of using conventional optics and increasing the ease of tunability, enabling highly efficient, arbitrary control over the pulse structure.
Fig.~\ref{fig:beam_approx}(a)-(c) shows the transverse profile at the input plane before a focusing optic for a negative radially chirped beam with radial polarization, along with a twelve beam and a four beam approximation of 1D spatially chirped, linearly polarized beams oriented as shown.
Each beam in the approximation has 1/N the energy of the radial beam to ensure that the energy of each configuration remains constant, where N is the number of beams.
The 1D spatially chirped beams are arranged such that they are equally spaced and have the same orientation, spatial chirp, $\delta r$, and polarization as the radially chirped beam.
The input aperture must be filled to better approximate the radial profile.
The aperture can be filled with two methods: by using more beams, as in Fig. \ref{fig:beam_approx}(b), or using fewer beams and increasing the un-chirped beam radius, shown in Fig. \ref{fig:beam_approx}(c), which used four beams whose radius was increased by a factor of 4.

Fig. \ref{fig:beam_approx}(d)-(f) shows the transverse dimensions of the focal plane for the radial, twelve, and four beam cases, respectively.
The intensities are normalized to the peak intensity of the radial beam.
The approximations reduce to a similar spot, where the twelve beam reaches 48\% of the peak intensity, and the four beam reaches 77\%.
The 12 beam is more spatially similar to the radial case, but the expanded four beam achieves higher peak intensities.
The four beam is symmetric around the axis, maintaining cartesian symmetry.
These configurations are not the optimum and the exploration of the effect different values of $\delta r$ and $\beta_{\text{BAR}}$ have on the required number of additional beams and expansion ratios is outside the scope of this manuscript.



There is a minor loss in peak intensity in the approximation, but it does not take into account the larger losses from the grating inefficiencies in the radially symmetric beams. 
The multi-beam approach can utilize high efficiency gratings, while also providing a more practical approach to their generation, as conventional optics are used to create the spatially chirped beam and control the relative orientation and polarization.
Individual focusing optics can be used for each beam, reducing the aperture required for large $\beta_{\text{BAR}}$ and $\delta r$ configurations.
This offers a distinct advantage over the configuration used to generate the radial chirp (Fig. \ref{fig:concentric_grating}), which requires advanced optics to create the variety of patterns shown in Fig. \ref{fig:chirp_polarizations}(c),(d).
Additionally, the ability to arbitrarily control the orientation, configuration, and polarization of each beam line enables the ability to explore a variety of different patterns and phase effects.




\section{Conclusion}
With the increased interest in spatio-temporal control of ultrashort pulses and vector beams, it is important to study how these complex pulses evolve as they propagate through their foci.
The radially chirped beams presented have many interesting tunable properties and focusing effects.
Similar to focused 1D spatially chirped beams, a temporal focusing effect is observed.
In the case of radial chirp, a symmetric pulse front is created which may help reduce non-reciprocal, PFT dependent effects.
The radial chirp enables a largely tunable on-axis centroid velocity that can be super-luminal or sub-luminal.
The annular nature of the spatial profile naturally extends to radial and azimuthal polarization states, offering focal volume structures with enhanced or minimized longitudinal fields.
While the generation of radially chirped beams is limited by the availability of high efficiency, polarization insensitive optics,
the ability to approximate the radially chirped beam using only four 1D spatially chirped beams can be readily accomplished with high efficiency optical components.
These approximations maintain all the properties of the radially chirped beams, while offering a more practical approach to the generation and the ability to control each beam independently, offering a large parameter space to explore and the prospect of complicated space-time foci for ultra-high intensity lasers.
\begin{backmatter}
\bmsection{Disclosures}
The authors declare no conflicts of interest.
\bmsection{Data availability}
Data underlying the results presented in this paper along with the code used to generate the figures are available in Ref. \cite{code}.
\bmsection{Supplemental document}
See Supplement 1 for supporting content.
\end{backmatter}

\bibliography{bibliography}






\end{document}


\maketitle
\begin{figure}[h!]
\centering\includegraphics[width=\linewidth]{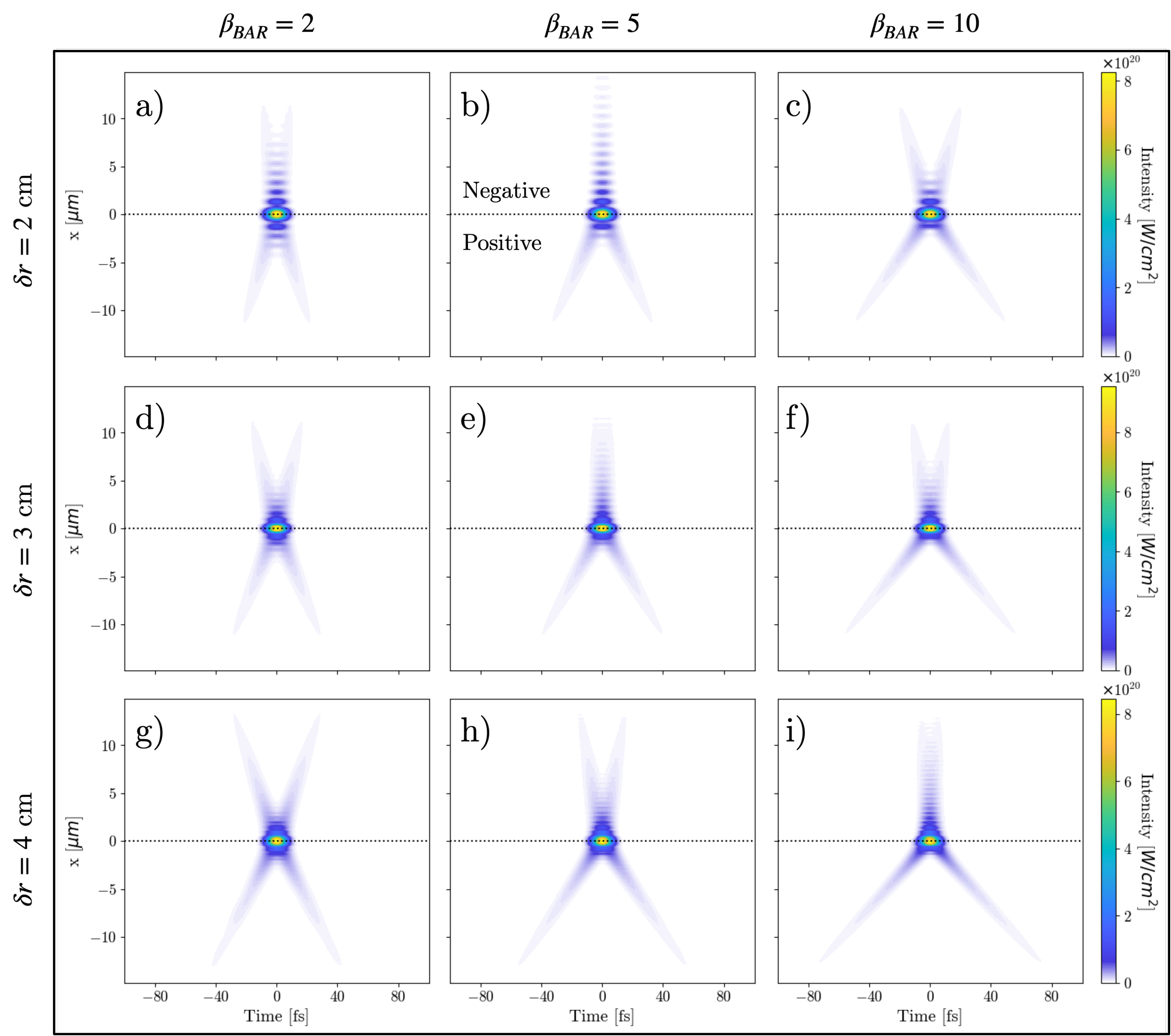}
\caption{(x,t) representations of the total intensity ($E_x^2 + E_y^2 + E_z^2$) of the focus ($f=5\; \text{cm})$ for a linearly polarized pulse spatially chirped with a $500$ lines/mm grating pair. The columns from left to right are pulses generated with a $\beta_{\text{BAR}}=2$, $\beta_{\text{BAR}}=5$, and $\beta_{\text{BAR}}=10$, respectively. The rows from top to bottom are pulses with a $\delta r=1$ cm, $\delta r=2$ cm, and $\delta r=3$ cm, respectively. Within each plot (a-i), negative spatial chirp is plotted in $x>0$ and positive spatial chirp is plotted in $x<0$ to provide direct visual comparison. A horizontal dotted line at $x=0$ in each plot indicates this separation.}
\label{fig:si_linear}
\end{figure}

\begin{figure}[h!]
\centering\includegraphics[width=\linewidth]{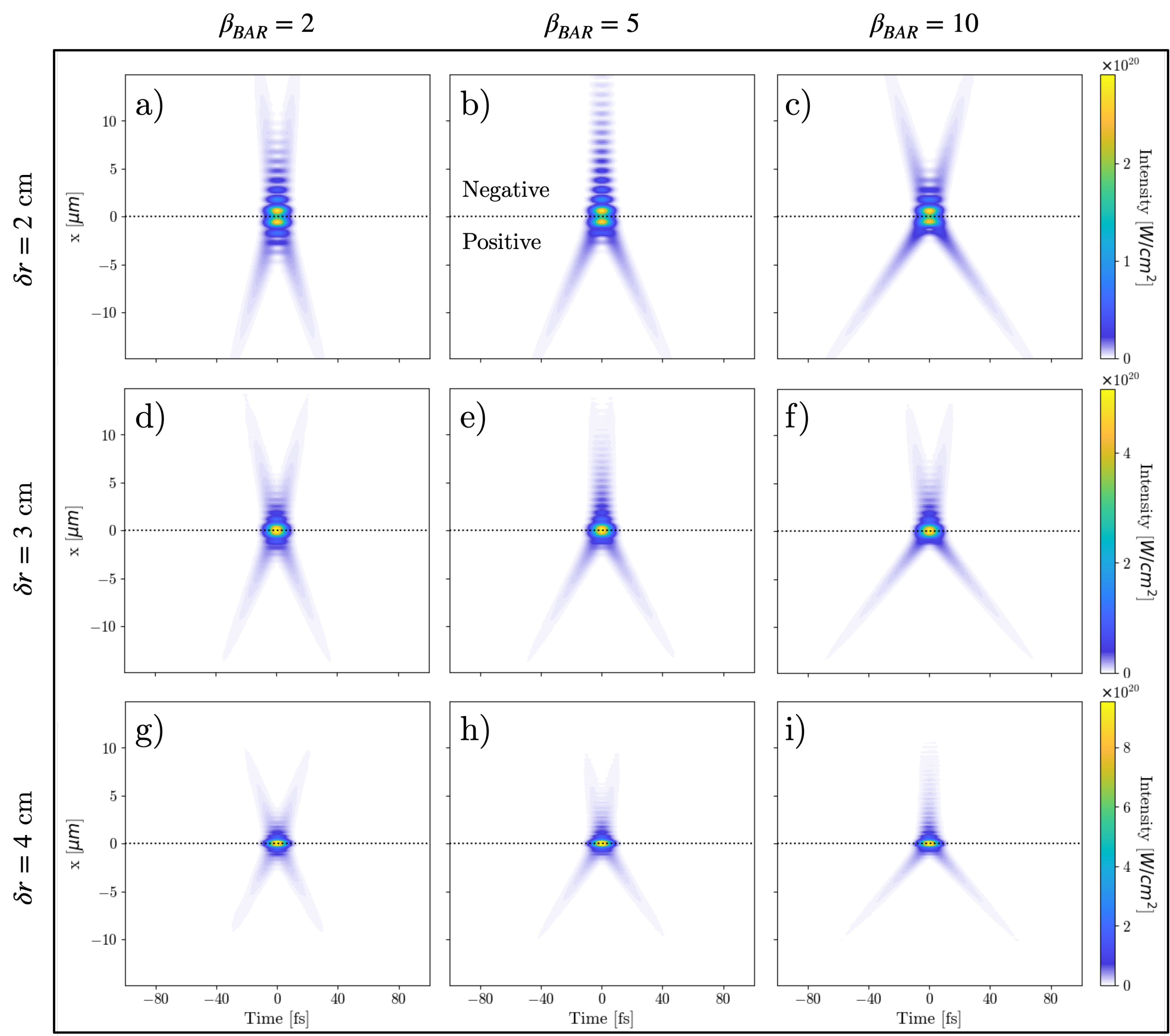}
\caption{(x,t) representations of the total intensity ($E_x^2 + E_y^2 + E_z^2$) of the focus ($f=5\; \text{cm})$ for a radially polarized pulse spatially chirped with a $500$ lines/mm grating pair. The columns from left to right are pulses generated with a $\beta_{\text{BAR}}=2$, $\beta_{\text{BAR}}=5$, and $\beta_{\text{BAR}}=10$, respectively. The rows from top to bottom are pulses with a $\delta r=1$ cm, $\delta r=2$ cm, and $\delta r=3$ cm, respectively. Within each plot (a-i), negative spatial chirp is plotted in $x>0$ and positive spatial chirp is plotted in $x<0$ to provide direct visual comparison. A horizontal dotted line at $x=0$ in each plot indicates this separation.}
\label{fig:si_linear}
\end{figure}

\begin{figure}[h!]
\centering\includegraphics[width=\linewidth]{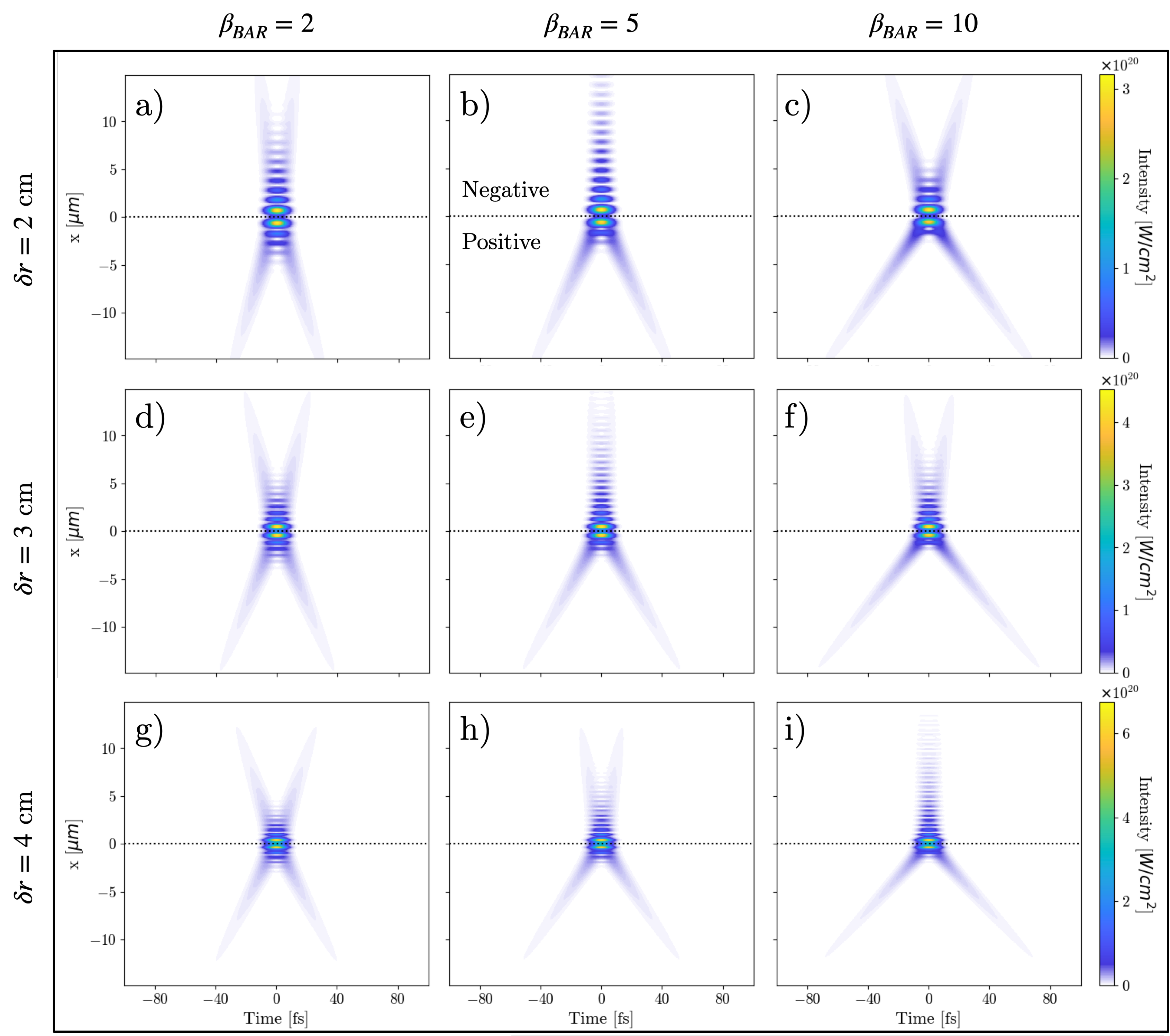}
\caption{(x,t) representations of the total intensity ($E_x^2 + E_y^2 + E_z^2$) of the focus ($f=5\; \text{cm})$ for an azimuthally polarized pulse spatially chirped with a $500$ lines/mm grating pair. The columns from left to right are pulses generated with a $\beta_{\text{BAR}}=2$, $\beta_{\text{BAR}}=5$, and $\beta_{\text{BAR}}=10$, respectively. The rows from top to bottom are pulses with a $\delta r=1$ cm, $\delta r=2$ cm, and $\delta r=3$ cm, respectively. Within each plot (a-i), negative spatial chirp is plotted in $x>0$ and positive spatial chirp is plotted in $x<0$ to provide direct visual comparison. A horizontal dotted line at $x=0$ in each plot indicates this separation.}
\label{fig:si_linear}
\end{figure}
